\documentclass[aps,prb,twocolumn,superscriptaddress,floatfix,longbibliography]{revtex4-1}

\usepackage{physics}
\usepackage{graphicx}
\usepackage{booktabs}
\usepackage{bm, bbm, amssymb, siunitx}

\newcounter{para}
\newcommand\mypara{\par\refstepcounter{para}\noindent \textbf{\thepara}\indent}

\usepackage{xcolor}
\definecolor{mblue}{RGB}{0, 118, 186}

\definecolor{mgreen}{RGB}{29, 177, 0}

\definecolor{mred}{RGB}{209, 28, 36}

\usepackage{hyperref}
\hypersetup{
    colorlinks=true,
    linkcolor=blue,     
    urlcolor=blue,
    citecolor=blue
}
\begin{document}

\title{van der Waals metamaterials}
    
\author{William Dorrell}
\author{Harris Pirie}
\affiliation{Department of Physics, Harvard University, Cambridge, MA, 02138, USA}
\author{S.~Minhal Gardezi}
\affiliation{School of Engineering and Applied Sciences, Harvard University, Cambridge, MA, 02138, USA}
\affiliation{Department of Physics, Wellesley College, Wellesley MA, 02481, USA}
\author{Nathan C. Drucker}
\affiliation{School of Engineering and Applied Sciences, Harvard University, Cambridge, MA, 02138, USA}
\author{Jennifer E. Hoffman}
\affiliation{Department of Physics, Harvard University, Cambridge, MA, 02138, USA}
\affiliation{School of Engineering and Applied Sciences, Harvard University, Cambridge, MA, 02138, USA}

\date{\today}

\begin{abstract}
Van der Waals heterostructures are an active frontier for discovering emergent phenomena in condensed matter systems. They are constructed by stacking elements of a large library of two-dimensional materials that couple together through van der Waals interactions. However, the number of possible combinations within this library is staggering, so fully exploring their potential is a daunting task.  Here we introduce van der Waals metamaterials to rapidly prototype and screen their quantum counterparts. These layered metamaterials are designed to reshape the flow of ultrasound to mimic electron motion. In particular, we show how to construct analogues of all stacking configurations of bilayer and trilayer graphene through the use of interlayer membranes that emulate van der Waals interactions. By changing the membrane's density and thickness, we can also reach coupling regimes far beyond that of naturally occurring graphene. We anticipate that van der Waals metamaterials can be used to explore, extend, and inform future electronic devices. Furthermore, they allow the transfer of useful electronic behavior to acoustic systems, such as flat bands in magic-angle twisted bilayer graphene, which may aid the development of super-resolution ultrasound imagers. 
\end{abstract}

\maketitle

\section{Introduction}

\mypara 
The recent excitement surrounding van der Waals (vdW) heterostructures stems from the diverse emergent phenomena that can arise by layering two-dimensional (2D) materials like graphene and other xenes, or transition metal dichalcogenides \cite{Ajayan2016Two-dimensionalMaterials,Novoselov20162DHeterostructures,Geim2013VanHeterostructures}. Such vdW heterostructures are poised to contribute to transformative technologies including ultra-thin low-energy transistors  \cite{Avsar2015Air-stableTransistors}, photodetectors \cite{Wang2013High-responsivityPhotodetectors}, and light-emitting diodes \cite{Withers2015WSe2Temperature}. Recently, their capabilities were expanded to include exotic many-body quantum behaviors such as unconventional superconductivity, which can occur in twisted bilayer graphene (TBG) \cite{Cao2018CorrelatedSuperlattices,Cao2018UnconventionalSuperlattices}.   However, finding the most interesting systems and behaviors is a time-consuming task, given the countless stacking combinations afforded by the ever-increasing library of 2D materials and tuning parameters such as twist angle. Twist angle is a particularly exciting but challenging new parameter, because the small `magic angle' of 1.08$^{\circ}$ that gives rise to such exotic effects in TBG also presents forbidding computational difficulties due to the increased unit cell size.  An open problem is to develop a method for rapidly predicting and prototyping vdW heterostructures to create a tight feedback loop for their technological advancement. 

\begin{figure}
  \includegraphics[width=\linewidth]{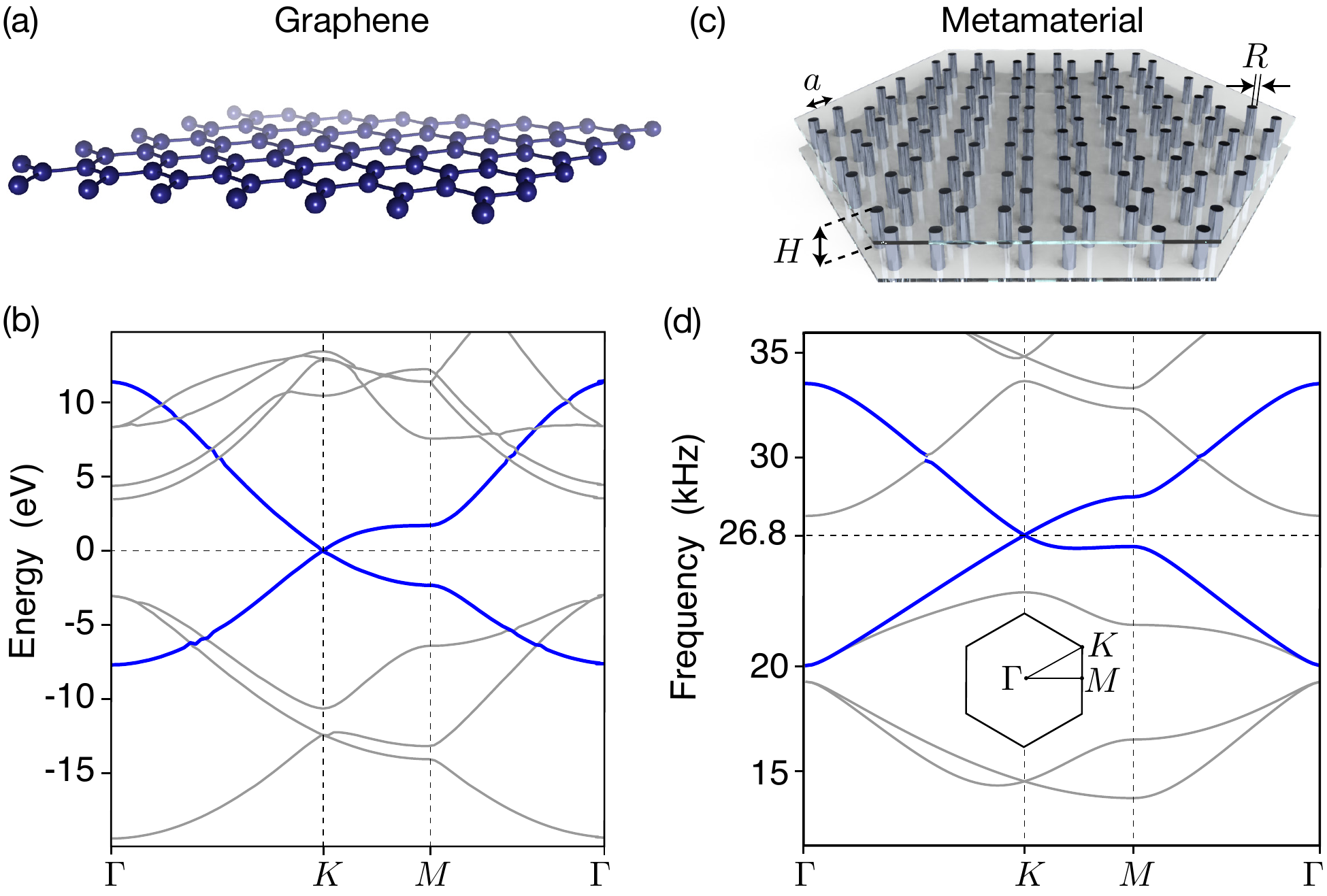}
  \caption{\textbf{Phononic metamaterial analogue of graphene.} (\textbf{a}) In graphene, the honeycomb lattice symmetry leads to (\textbf{b}) a band structure featuring a linearly dispersive $K$-point Dirac cone (blue), adapted from Ref.~[\onlinecite{KN12}]. (\textbf{c}) We designed a phononic metamaterial to imitate graphene by arranging a honeycomb lattice of steel pillars in air. The pillars have radius $R=0.32$ cm, height $H=0.1$ cm, and spacing $a=1$ cm. (\textbf{d}) Their simulated phononic band structure recreates the Dirac cone in graphene (blue).}
  \label{fig:1}
\end{figure}

\mypara 
In the last few years, phononic metamaterials have emerged as a promising platform for mimicking condensed matter systems \cite{TS12, Torrent2013ElasticPlates, SusstrunkScience2015, NashProcNatlAcadSci2015, He2016AcousticTransport}.  They are appealing quantum analogues because their governing wave equations are straightforward, making calculations fast; they can be quickly fabricated; and their properties derive from macroscopic structures with continuously tunable geometry, as opposed to the limited discrete elements of the periodic table. A carefully constructed phononic metamaterial can host propagating sound waves that closely resemble the behavior of electrons moving in solids. For example, the Dirac-like electronic bands of graphene, which rely on the $C_6$ symmetry of its honeycomb carbon lattice (Fig.~\hyperref[fig:1]{1(a-b)}), can be reproduced in the phonon band structure of a honeycomb arrangement of steel pillars in air (Fig.~\hyperref[fig:1]{1(c-d)}). This general framework has been applied to yield analogues of graphene in longitudinal phononics \cite{Lu2014DiracWaves,Mei2012First-principlesCrystals}, surface acoustics \cite{Yu2016SurfaceGraphene}, photonics \cite{Ochiai2009PhotonicStates}, and mechanics \cite{Torrent2013ElasticPlates,Kariyado2015ManipulationGraphene}. Given these successes, it is natural to ask whether the same strategy works for multi-layer systems.
Existing metamaterial designs have made innovative use of multi-layer structures \cite{Chiang2011AMetamaterials,Wang2019AMetamaterials}, including recent work that coupled two-layered sonic crystals to create topological phases \cite{JCW18}. These findings represents important steps towards mimicking multi-layer heterostructures. However, without control of the strength of interlayer coupling, accurately recreating the diverse library of vdW heterostructures remains an elusive goal. 

\mypara
Here we demonstrate a versatile framework for mimicking layered vdW heterostructures with acoustic metamaterials. Specifically, we numerically explore the coupling strength of an interlayer membrane as a function of material properties using the commercial finite-element modeling software {\sc comsol multiphysics}. By tuning the membrane properties, we discover a simple phenomenological model for metamaterial interlayer coupling that allows us to recreate all stacking configurations of bilayer and trilayer graphene, and furthermore to reach coupling regimes far beyond those in naturally occurring graphene. Our work opens a different path to explore vdW heterostructures, which could uncover new phenomena and inform the fabrication of future quantum materials. In the opposite direction, our work guides the translation of electronic vdW phenomena to phononic systems, which could stimulate useful acoustic devices.

\begin{figure}
  \includegraphics[width=\linewidth]{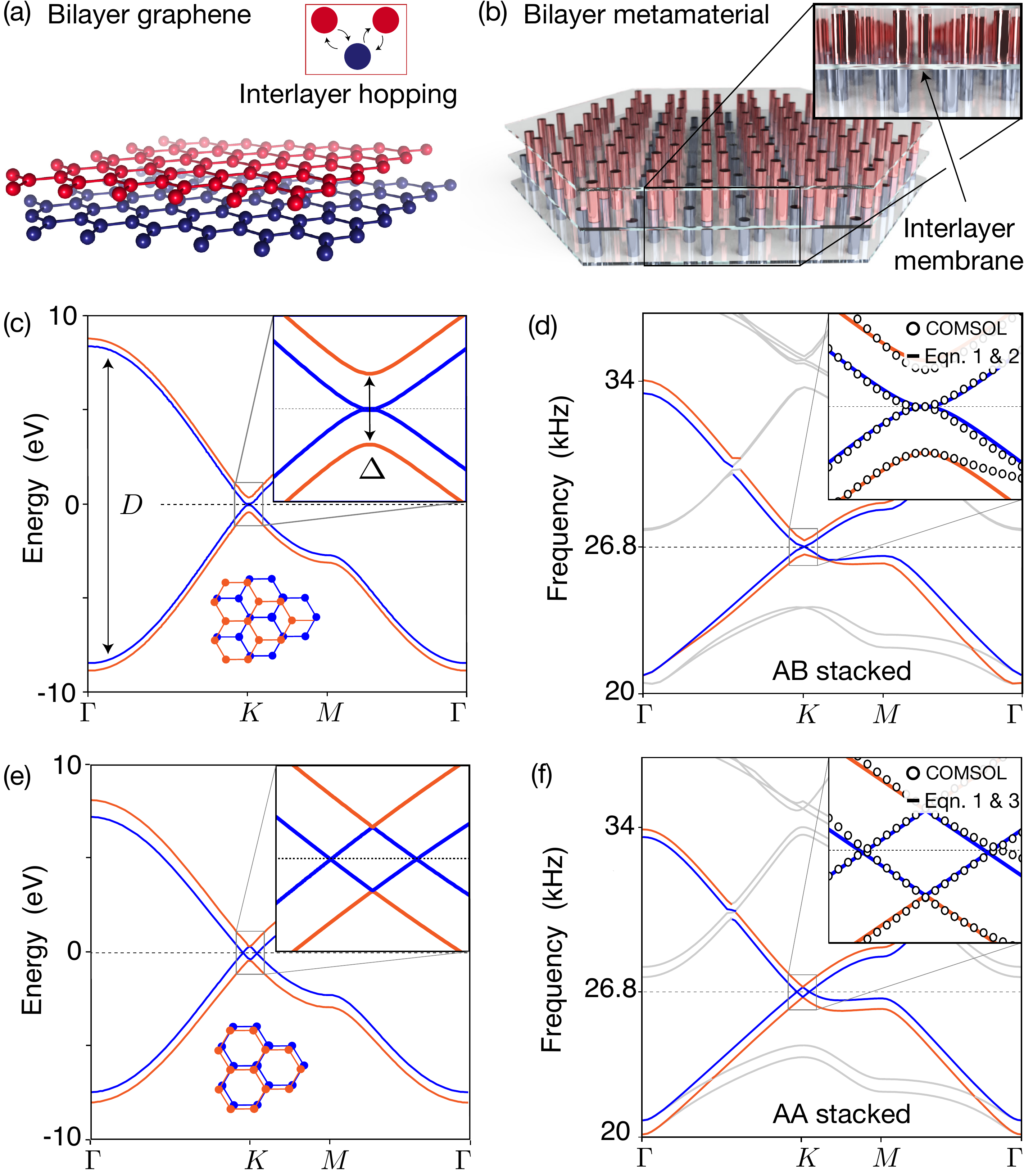}
  \caption{
  \textbf{Phononic metamaterial analogue of bilayer graphene.} (\textbf{a}) We recreated  interlayer hopping in bilayer graphene using (\textbf{b}) two layers of honeycomb metamaterial separated by a flexible membrane. (\textbf{c}) In AB-stacked bilayer graphene, the Dirac cones hybridize to create a `kissing' band structure. (\textbf{d}) With an interlayer membrane made of $0.19$-mm thick HDPE, the same effect is seen in our bilayer metamaterial. This effect is well described by the same tight-binding Hamiltonian used to describe bilayer graphene (inset). (\textbf{e}) In AA-stacked bilayer graphene, the structure of the Dirac cone changes. (\textbf{f}) By stacking our metamaterial in the AA configuration, it accurately captures the AA graphene Dirac cone structure. Panels (c) and (e) are tight-binding calculations reprinted with minimal alterations from  Ref.~[\onlinecite{Rozhkov2016ElectronicSystems}], with permission from Elsevier.}
  \label{fig:2}
\end{figure}

\section{Phononic bilayer graphene}

\mypara
Building on previous work, we started from a monolayer phononic metamaterial designed to emulate the Dirac cone in graphene \cite{ He2016AcousticTransport,Chen2014AccidentalCrystal,Pirie2018TopologicalLogic}. The presence of a Dirac cone relies only on the $C_6$ symmetry of the unit cell, but its shape can be tuned by material choice and pillar size.  For convenience, our device consists of a honeycomb lattice of steel pillars in air, as shown in Fig.~\hyperref[fig:1]{1(c)}. We calculated its band structure for various configurations of pillar radius $(R)$, height $(H)$, and spacing $(a)$, and found a close match to graphene when $R = 0.32$ cm, $H = 0.1$ cm, and $a = 1$ cm (Fig.~\hyperref[fig:1]{1(d)}). Importantly, our metamaterial contains an isolated Dirac cone like graphene, despite small differences in the surrounding bands. Consequently, for frequencies close to 26.8 kHz our acoustic device responds similarly to undoped graphene.

\mypara
Our first advance is to demonstrate a phononic analogue of bilayer graphene by stacking two honeycomb metamaterials on top of each other. In its natural state, bilayer graphene stacks in an AB configuration (Fig.~\hyperref[fig:2]{2(a)}). Each layer contributes an identical Dirac cone, $E(\mathbf{k})$, to the band structure. The two Dirac cones then couple through interlayer hopping, $\Delta$, to generate a characteristic band structure whereby the linear Dirac cones are replaced by parabolic `kissing' bands \cite{Rozhkov2016ElectronicSystems} (Fig.~\hyperref[fig:2]{2(c)}). The combined structure can be described by a simple tight-binding Hamiltonian \cite{Rozhkov2016ElectronicSystems,McCann2013TheGraphene}, 
\begin{align}
    \label{eqn:tbmodel}
    \mathcal{H}(\mathbf{k}) &= \begin{bmatrix}
    E(\mathbf{k}) & \delta \\
    \delta^T & E(\mathbf{k})
    \end{bmatrix},
    \text{ where } 
    E(\mathbf{k}) = \begin{bmatrix}
    0 & v_\mathrm{F}\mathbf{k} \\
    v_\mathrm{F}\mathbf{k} & 0
    \end{bmatrix},
\end{align}
\begin{align}
    \label{eqn:dab}
    \delta &= \delta_{AB} = \frac{1}{2}\begin{bmatrix}
    0 & 0 \\
    \Delta & 0
    \end{bmatrix},
\end{align} 
and $v_\mathrm{F}$ is the Fermi velocity. In this model, $E(\mathbf{k})$ describes hopping between sublattice sites within a single layer, while $\delta_{AB}$ describes the first-order interlayer coupling, which always occurs between inequivalent sub-lattice sites and so contributes an off-diagonal term.  To compare the magnitude of graphene's interlayer hopping to that in our metamaterial, we used a dimensionless coupling metric, $\widetilde{\Delta} = \Delta/D$, where $D$ is the Dirac-cone bandwidth. In bilayer graphene, $\widetilde{\Delta} = 4.7\% $ based on tight-binding fits to experimental data and {\it ab initio} calculations (see Fig.~\hyperref[fig:2]{2(c)} and Ref.~[\onlinecite{Rozhkov2016ElectronicSystems}]).

\mypara 
Our metamaterial analogue of bilayer graphene controllably couples phonons between layers through the use of an intermediary membrane (Fig.~\hyperref[fig:2]{2(b)}). Intuitively, as phonons in one layer propagate, they induce matching oscillations in the membrane, which links to the other layer and promotes the desired layer-to-layer crosstalk. To achieve a similar coupling magnitude as in bilayer graphene, we used a 0.19-mm thick sheet of high-density polyethylene (HDPE) as the intermediary layer. We computed the phononic band structure of our bilayer metamaterial using {\sc comsol multiphysics} (open circles in Fig.~\hyperref[fig:2]{2(d)}). All simulations contain additional 0.19-mm thick HDPE sheets on the outer edges to confine the modes to the 2D layers. Strikingly, we found `kissing' bands in the bilayer metamaterial, analogous to those in bilayer graphene. Furthermore, the phononic bands are well described by the same tight-binding model (Eqn.~\ref{eqn:tbmodel} and \ref{eqn:dab}, solid lines in Fig.~\hyperref[fig:2]{2(d)}) with reasonable values for the sound velocity $v_s=140$ m/s (analogous to $v_\mathrm{F}$ in graphene) and phonon bandwidth $\Delta = 700$ Hz, yielding $\widetilde{\Delta}=4.7\%$. Thus, the intermediary HDPE sheet quantitatively reproduces the interlayer hopping in bilayer graphene. 

\mypara
The first key advantage of vdW metamaterials is the versatility of their stacking configuration. In bilayer graphene, stacking arrangements other than AB  host a rich range of physical phenomena \cite{DeAndres2008StrongLayers, Cao2018CorrelatedSuperlattices}. Yet non-AB stacking is technically challenging to fabricate in graphene, which tends to relax locally back to AB, making it especially difficult to isolate and empirically confirm a particular uniform configuration. On the other hand, constructing translated or twisted metamaterials presents no additional difficulty, allowing analogues of quantum electronic phenomena to be rapidly and unambiguously explored. To illustrate this point, we recreated the band structure of AA-stacked bilayer graphene by translating our AB-stacked metamaterial, but retaining the same HDPE interlayer (Fig.~\hyperref[fig:2]{2(e-f)}). The band structure is well described by the tight-binding Hamiltonian in Eqn.~\ref{eqn:tbmodel}, but with a modified coupling matrix that instead encodes interlayer hopping between vertically-aligned, identical sub-lattice sites: 
\begin{align}
    \label{eqn:daa}
    \delta = \delta_{AA} = \frac{1}{2}
    \begin{bmatrix}
    \Delta & 0 \\
    0 & \Delta
    \end{bmatrix}.
\end{align}
This model accurately describes both the phononic and electronic AA-stacked systems, as shown in the insets to Fig.~\hyperref[fig:2]{2(e-f)}. In addition, it captures the behavior of both AA-stacked and AB-stacked metamaterials using the same values of $v_\mathrm{F}$ and $\Delta$, simply by switching the $\delta$ matrix.  

\begin{figure} [t]
  \includegraphics[width=0.95\linewidth]{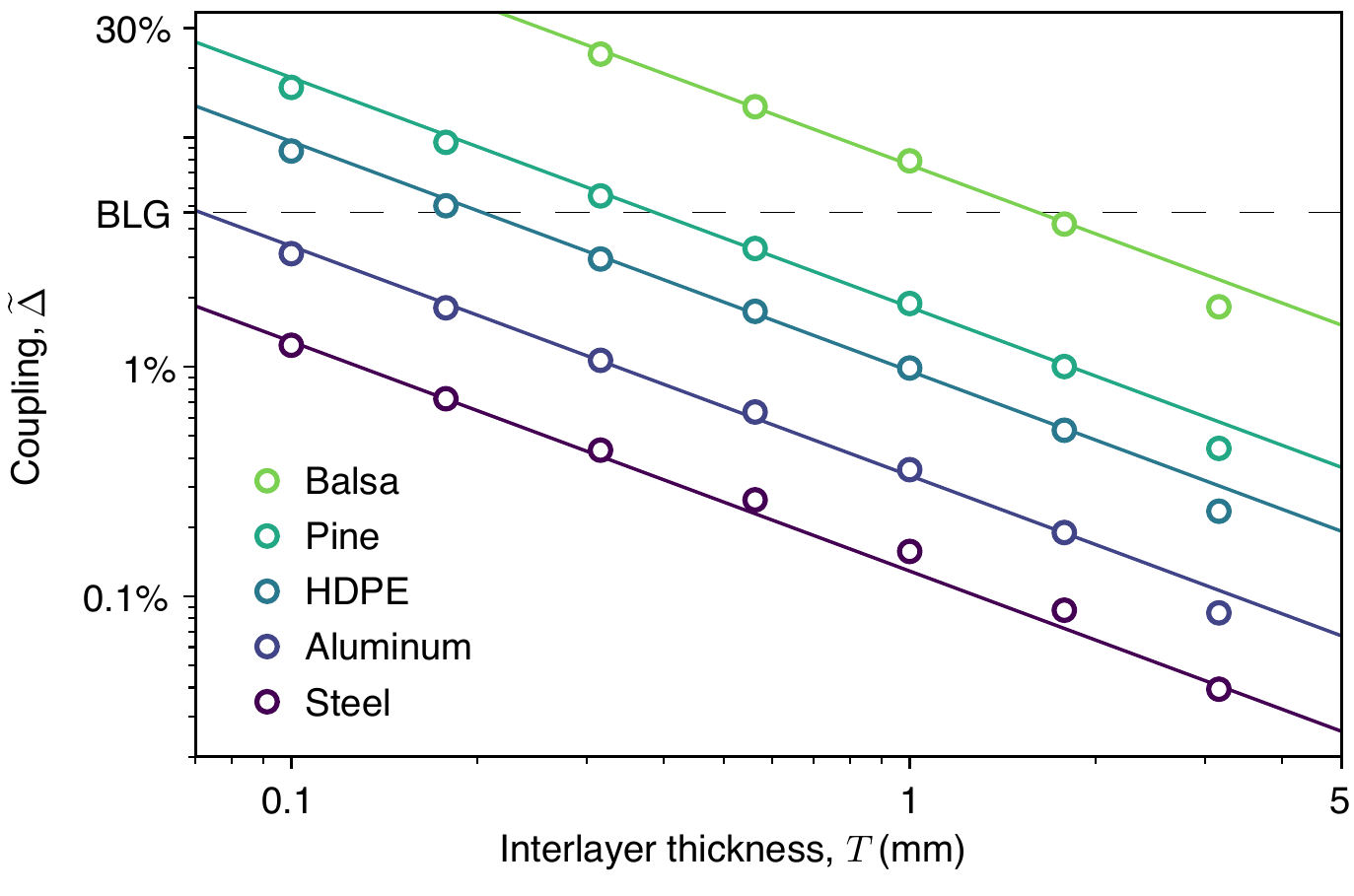}
  \centering
  \caption{
  \textbf{Interlayer coupling is controllable over two orders of magnitude.}
  The computed dimensionless coupling strength ($\widetilde{\Delta}$, open circles) varies inversely with interlayer density ($\rho$) and thickness ($T$):  $\widetilde{\Delta} = \Delta/D \propto  1/(\rho T)$ (solid lines), where $\Delta$ is the interlayer hopping and $D$ is the bandwidth. With common materials, it is possible to engineer interlayer coupling to be an order of magnitude larger or smaller than in natural bilayer graphene (4.7\%, dashed line). }
  \label{fig:3}
\end{figure}

\mypara
The second key advantage of vdW metamaterials is the ease with which membrane properties can be changed to explore diverse coupling regimes. In electronic vdW heterostructures, the interlayer coupling is often seen as a fixed property, and experiments that vary it are technically challenging. For instance, a substantial vertical pressure of 1.3 GPa is required to effect just a 20\% increase in the interlayer coupling strength of TBG by pushing the layers closer together \cite{Yankowitz2019TuningGraphene}. Yet, this additional tuning knob has permitted some remarkable discoveries; for example, pressure increases the magic angle and the superconducting transition temperature in TBG \cite{Yankowitz2019TuningGraphene, CFJ18}. In our system, tuning the interlayer coupling is as simple as using a thicker membrane or changing its composition. Intuitively, the same pressure variation causes a flexible membrane to move more than a stiff membrane, so the interlayer coupling increases as the membrane becomes thinner or less dense. We computed the band structure of our AB-stacked bilayer metamaterial to quantify $\widetilde{\Delta}$ as a function of thickness for five different membrane materials, as shown in Fig.~\ref{fig:3}. We found that $\widetilde{\Delta}$ was largely independent of the interlayer membrane's speed of sound over a wide range of values. For example, an artificial order-of-magnitude increase in the speed of sound of the HDPE membrane in Fig.~\hyperref[fig:2]{2(b)}, without altering its density ($\rho$) or thickness ($T$), produced only a minimal change in $\widetilde{\Delta}$ from 4.7\% to 4.6\%. To a good approximation, $\widetilde{\Delta}$ varies inversely with $\rho$ and $T$ following the phenomenological rule:
\begin{align}
     \widetilde{\Delta} \equiv \frac{\Delta}{D} \propto  \frac{1}{\rho T}.
\end{align}
These strong dependencies make it possible to vary $\widetilde{\Delta}$ over more than two orders of magnitude using only common household materials, like steel or wood.  Thus, candidate twistronic materials can be categorized based on their intralayer and interlayer coupling strengths, then prototyped as simple phononic metamaterials by adjusting the pillar and membrane properties. In addition, the interlayer thickness could be spatially textured to capture variations in coupling strength due to the inherent lattice corrugation or relaxation in many vdW heterostructures, including twisted bilayer graphene \cite{Novoselov20162DHeterostructures,WSK15, YEC19, TKV19}.

\section{Towards Full vdW Heterostructures}

\begin{figure}[t!]
  \includegraphics[width=\linewidth]{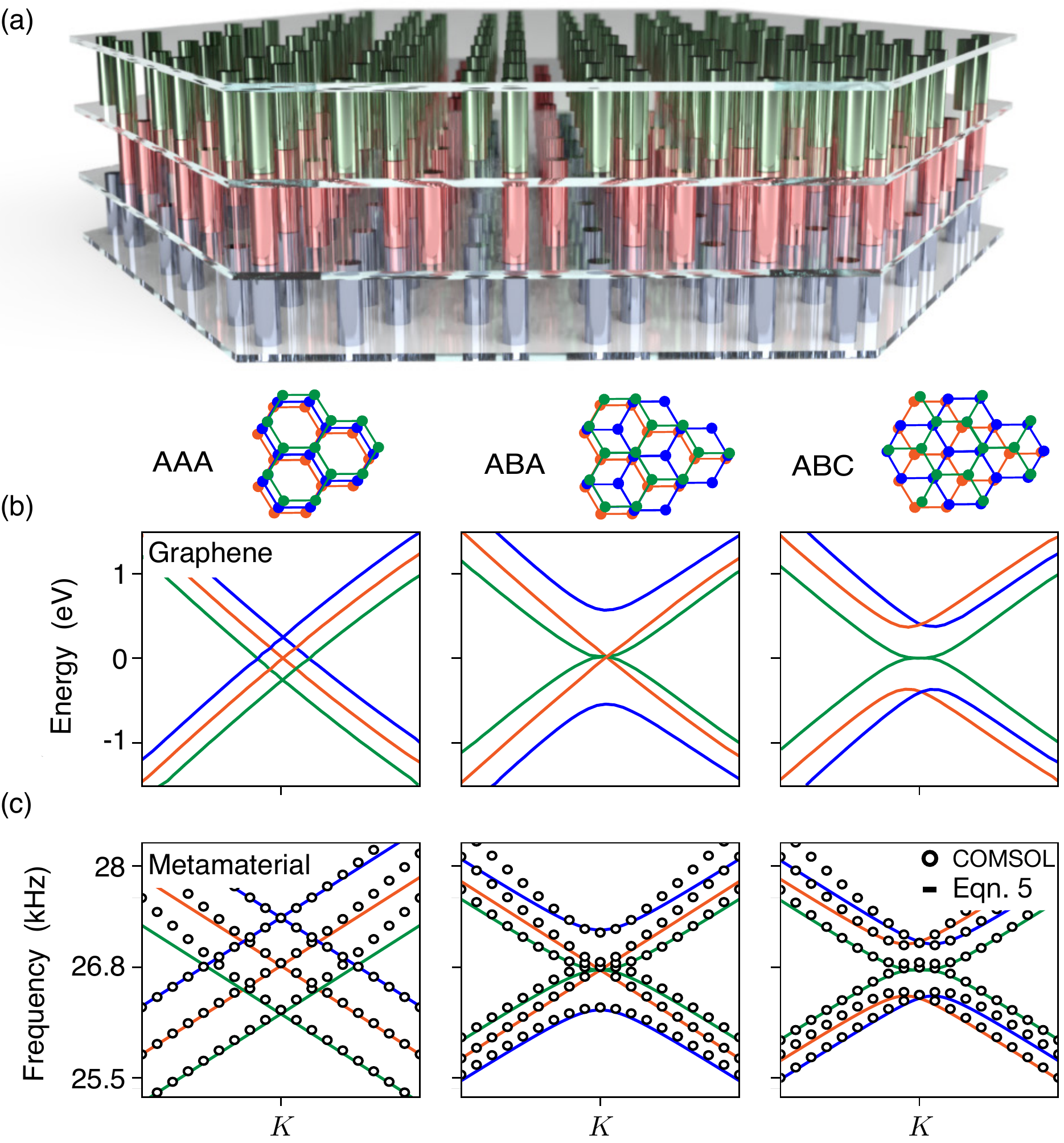}
  \caption{
  	\textbf{Phononic metamaterial analogue of trilayer graphene}. 
	{\bf (a)}  Our trilayer phononic metamaterial comprises three honeycomb lattices of steel pillars separated by $0.19$-mm thick HDPE layers. 
	\textbf{(b)} The Dirac cone structure of trilayer graphene displays qualitatively different behavior in each of its three different stacking configurations, adapted with permission from Ref.~[\onlinecite{Bao2017Stacking-DependentSpectroscopy}]. Copyright 2017, American Chemical Society. 
	\textbf{(c)} Each behavior is recreated in a trilayer metamaterial by matching the stacking pattern. The calculated band structure (open circles) is well described by a simple two-parameter Hamiltonian (Eqn.~\ref{eqn:trilayer}, solid lines).}
  \label{fig:4}
\end{figure}

\mypara 
Our interlayer coupling scheme generalizes beyond bilayer metamaterials. Quantum vdW heterostructures combine elements from a vast library of 2D materials to realize emergent behavior that further diversifies as the number of layers increases.  To demonstrate this trend in vdW metamaterials, we simulated an analogue of trilayer graphene by stacking three honeycomb lattices of steel pillars, interspersed with the same HDPE membranes used previously (Fig.~\hyperref[fig:4]{4(a)}). Trilayer graphene has three possible stacking configurations, each encoded by a different interlayer hopping matrix, $\delta$.  In each case, we engineered a similarly-stacked trilayer metamaterial with interlayer interactions that promote a nearly identical band structure to trilayer graphene, as shown in Fig.~\hyperref[fig:4]{4(b-c)}.  Both the electronic and phononic system can be described by the same tight-binding Hamiltonian, which is a simple extension of Eqn.~\ref{eqn:tbmodel},
\begin{align}
\label{eqn:trilayer}
\mathcal{H}(\mathbf{k}) = \begin{bmatrix}
    E(\mathbf{k}) & \delta_1 & 0\\
    \delta_1^T & E(\mathbf{k}) & \delta_2\\
    0 & \delta_2^T & E(\mathbf{k})
    \end{bmatrix},
\end{align} 
where the stacking-dependent $\delta$ matrices are 
\begin{align}
 \delta_{AAA}^{(1)} &= \delta_{AAA}^{(2)} = \frac{\delta_{AA}}{\sqrt{2}},
    \\
 \delta_{ABA}^{(1)} &= \frac{\delta_{AB}}{\sqrt{2}}, \qquad \delta_{ABA}^{(2)} = \frac{\delta^T_{AB}}{\sqrt{2}},
    \\
\delta_{ABC}^{(1)} &= \delta_{ABC}^{(2)} = \frac{\delta_{AB}}{\sqrt{2}}.
\end{align} 

\mypara 
Unlike their quantum counterparts, vdW metamaterials can be tuned continuously to span a large space of band structures. Although we have focused primarily on analogues of graphene, parallels to other vdW materials can be easily drawn by tuning the geometry and composite materials of the metamaterial. For example, hexagonal boron nitride (hBN) is a common ingredient in many vdW heterostructures because it hosts a large insulating band gap \cite{Lee2011ElectronNitride}. It consists of two interpenetrating triangular lattices of boron and nitrogen, and consequently breaks the $C_6$ symmetry of the unit cell to open a gap at the $K$ point.  Similarly, an analogue of hBN can be produced by breaking the $C_6$ symmetry of the graphene metamaterial in Fig.~\hyperref[fig:1]{1(c)}, for example by using pillars of different radii. Unlike the electronic version, the resulting band gap in the metamaterial can be continuously tuned because the radii may be chosen arbitrarily. Single-layer metamaterials exhibiting such wide band gaps have been extensively explored in the literature \cite{Jia2018DesigningGaps}. Our coupling scheme allows them to be incorporated into vdW metamaterials, affording enormous flexibility in recreating, exploring, and extending the emergent electronic phenomena of quantum vdW heterostructures.

\section{Conclusions}

\mypara 
The essential ingredient of vdW heterostructures---their interlayer coupling---extends beyond electronic systems.  By adding a flexible membrane between phononic metamaterials, we demonstrated that this coupling can be accurately recreated in acoustics.  The resultant vdW metamaterials permit the rapid exploration of diverse stacking combinations and extensive coupling regimes that are challenging to reach in electronic materials.  To illustrate their potential, we developed macroscopic acoustic analogues of every stacking configuration of bilayer and trilayer graphene. Our results provide a guide for designer vdW metamaterials, which may focus and inform future vdW-based electronic devices. Conversely, they permit mapping intriguing electronic phenomena to phononics, like the flat bands in magic-angle TBG, which may provide a new platform for non-linear acoustics.  Finally, our design scheme applies directly to photonic metamaterials through a simple mapping of variables \cite{Mei2012First-principlesCrystals}.

\bibliography{refs.bib}

\end{document}